# Three-Dimensional Lagrangian Coherent Structures in Patients with Aortic Regurgitation

Wissam Abdallah[1], Ahmed Darwish[1,2,*], Julio Garcia[3,4] and Lyes Kadem[1]

[1] Laboratory of Cardiovascular Fluid Dynamics, Concordia University, Montréal, QC H3G 1M8, Canada

[2] Mechanical Power Engineering Department, Assiut University, 71515, Egypt

[3] Department of Cardiac Sciences, University of Calgary, Calgary, AB, T2N 2T9, Canada

[4] Department of Radiology, University of Calgary, Calgary, AB, T2N 2T9, Canada

(*) Corresponding author – email : ahmeddiaa@aun.edu.eg

# Abstract

Understanding blood transport in cardiovascular flows is important for managing patients with cardiovascular disease. In this study, three-dimensional Lagrangian coherent structures have been extracted for the first time in both healthy patients and patients with aortic regurgitation. To achieve this, a computationally efficient approach based on Lagrangian descriptors was employed with four dimensional (4D) magnetic resonance imaging velocity fields. In healthy subjects, Lagrangian coherent structures (LCSs) analysis revealed well-defined mitral jet structures during early filling, directing flow toward ejection during systole. For patients with aortic regurgitation, complex flow structures included interactions between the mitral and regurgitant jets, indicating altered blood transport mechanisms. This study highlights the ability of Lagrnagian descriptors to extract coherent structures from patient-specific 4D flow MRI data in a computationally efficient way. It also underscores the importance of extracting three-dimensional Lagrangian coherent structures to gain a better understanding of the complex interaction between the mitral inflow and the regurgitant jet.

Enhancing our understanding of blood mixing, segregation, and stasis within cardiovascular flows can be facilitated by uncovering the underlying mechanism of blood transport [1]. To achieve this goal, extracting Lagrangian coherent structures (LCSs) has become a common approach to reveal the blood transport mechanism in cardiovascular flows [1–12]. LCSs are fluid transport barriers that organize the transported fluid into relatively separate regions with different dynamics. A deeper look into LCSs (from the study of dynamical systems) reveals them as stable and unstable manifolds distributed along the transported fluid, governing how the fluid is transported. For instance, vortical structures can be defined by extracting their boundaries using LCS, as demonstrated in [2,3,5,6,13–16].

The necessity to extract 3D LCSs has been highlighted by Sulman et al. [17] where they found that for sufficiently large vertical shear, full 3D calculation becomes necessary. Zhang et al. [18] extracted LCSs in fluid flow in dynamic stall of a pitching airfoil where they were able to track the evolution and motion of primary leading-edge vortex and trailing edge vortex. Cao et al. [19] introduced LCSs method to forced convection heat transfer around a circular cylinder in laminar flow regime where they related the instantaneous thermal and vorticity patterns, lift coefficient, drag coefficient and Nusselt number to the extracted LCSs. Aksamit [20] introduced a mapping technique of the Cauchy-Green tensor eigenvalues in order to quantify and visualize the three-dimensional LCSs. Bian et al. [21] proposed integrating a discrete phase model (DPM) along with a CFD solver to efficiently extract 3D LCSs (using FTLE).

In cardiovascular flows, based on experimental or CFD data, LCSs have been used to uncover structures that control drug transport and optimize it for applications such as targeted drug delivery [14]. Furthermore, LCSs have been instrumental in characterizing the link between intraventricular flow and thromboembolic risk [4], understanding the mechanism behind aortic aneurysm growth by revealing flow stagnation volumes [7], assessing the impact of flow diverters on intracranial aneurysm flow [22,23], identifying the hemodynamic effects of left superior pulmonary vein resection in the left atrium[24], revealing the interventricular fluid transport topology from Doppler echocardiographic velocity fields[25]. Blood transport downstream of a calcified aortic valve was also investigated using LCS [26]. In vivo, utilizing 4D flow MRI, LCSs have been employed to characterize the interventricular vortex ring volume during diastole [5] and to analyze intraventricular kinetic energy in patients with pulmonary regurgitation [27]. For a comprehensive review of the clinical assessment of interventricular flow by extracting LCSs, the reader is referred to Ref.[28].

With the increasing interest in extracting LCS from cardiovascular flows using 4D PC-MRI, it is crucial to address the major challenges that hinder their extraction from 4D flows. The most common approach for extracting LCS is a geometrical approach that relies on evaluating the finite-time Lyapunov exponents (FTLEs). Central to this approach is finding the largest eigenvalue of the right Cauchy-Green tensor (CGT), which indicates the maximum amount of stretching at a given initial point in a fluid domain. Often, the right CGT is computed for a dense grid of particles with their trajectories being traced along a finite time interval. Each grid point corresponds to a Lagrangian trajectory. By considering an infinitesimally small perturbation **δ** (equal to the grid spacing) between the trajectories, the material curves (for 2D flows) or surfaces (for 3D flows) which govern the fluid transport can be extracted by tracking the evolution of **δ**(t) [19], [20]. Now, the perturbation that is aligned with the dominant eigenvector of the right CGT will experience the largest stretching, while the perturbation that is aligned with the weakest eigenvector will witness the largest compression. Then, the average growth exponents derived from the evolution of these singular values are called the finite-time Lyapunov exponents (FTLEs). Notably, for 2D flows, we have to evaluate a 2x2 CGT at each grid point. This becomes more complex for 3D flows, where a 3x3 CGT is evaluated at each grid point. To overcome the computational cost associated with FTLE, an alternative approach that does not involve such intensive computations is needed. This can be achieved using a trajectory-based approach like Lagrangian Descriptors (LD). LD only requires the evaluation of a scalar physical property (e.g., velocity, vorticity) along a set of Lagrangian trajectories, with LCS corresponding to abrupt changes in chosen property maps. LD-based LCS extraction employs an integral formulation and does not rely on velocity gradient computations like FTLE, reducing computational costs significantly, by at least five times for 2D flows [29]. Moreover, LD is more robust in the presence of experimental measurement uncertainties. LD along with a new global dynamics and chaos indicator was used to reveal chaotic regions in discrete and continuous dynamical systems[30]. García-Garrido et al. [31] applied LD to extract LCSs from three-dimensional velocity fields. Three-dimensional LD have been also used to reveal the spread of SARS-CoV-2 in a hospital isolation room from CFD results [32].

Therefore, this study aims to use Lagrangian Descriptors to extract, for the first time, three-dimensional Lagrangian coherent structures in healthy patients and patients with aortic regurgitation, a valvular pathology resulting in left ventricle overload due to ventricular filling from both the mitral valve and aortic valve.

The study cohort comprises eight healthy subjects and three patients with severe aortic regurgitation. All subjects were recruited from the observational Cardiovascular Imaging Registry of Calgary (CIROC). The University of Calgary Research Ethics Board approved the study, and all subjects provided written consent at the time of the scan. Research activities were conducted following the Declaration of Helsinki, and informed consents were recorded using dedicated software (CardioDITM, Cohesic Inc, Calgary, AB, Canada). For inclusion and exclusion criteria, please refer to [33].

CMR image acquisition involved a standardized cardiac imaging protocol for valvular diseases on all participants, utilizing a 3T MRI scanner (Siemens, Erlangen, Germany), following current published recommendations [34]. Imaging of the entire heart was achieved through standard electrocardiographic (ECG) gating, balanced time-resolved steady-state free precision (SSFP) cine imaging in short-axis, 3-chamber, 2-chamber, and 4-chamber views. Additionally, a 3D contrast-enhanced magnetic resonance angiography (CEMRA) of the whole heart was performed by administering 0.2 mmol/kg of gadolinium contrast (Gadovist®, Bayer Inc, Mississauga, ON, Canada).

For volumetric blood flow assessment, a time-resolved 3D phase-contrast MRI with three-directional velocity encoding was obtained using a 4D-flow MRI WIP from Siemens (WIP 845A). The entire heart was covered using sagittal slices. The acquisition was performed for 5 - 10 minutes followed by the administration of the contrast agent, using retrospective ECG gating and free-breathing supported by a diaphragmatic motion navigator. The following parameters were used for the 4D-flow acquisition: bandwidth = 455-495 Hz/Pixel, pulse repetition time = 4.53 – 5.07 ms, echo time = 2.01 – 2.35 ms, flip angle = 15 degrees, spatial resolution = 2.0 – 3.6 × 2.0 – 3.0 × 2.5 – 3.5 mm3, Venc = 150 – 400 cm/s, phases = 30, and temporal resolution = 25 – 35 ms. Total acquisition time varied depending on the patient's heart rate and respiratory navigator efficiency.

For 4D flow image analysis, 4D flow data were processed using EnSight software (CEI Inc., Apex, NC, USA) using 25 frames representing the 0.9-1s physiological heart cycle. The first step involved locating the region of interest by placing a plane cutting through the LV and simultaneously passing through both the mitral and the aortic valves. The most representative plane and time frame were selected to describe the diastolic and systolic trans-mitral and aortic flows. The second step was the segmentation of the LV from the total heart using a customized MATLAB library, allowing the generation of velocity vectors. Pathlines assessment facilitated understanding the visual and macroscopic effect of the AR on the LV and mitral valve.

Considering the three-dimensional velocity field **u** (**x**, *t*) at any given time instant t along the cardiac cycle, we placed uniformly spaced passive particles over a refined grid (**x**₀) at an initial time $t_0$. As each particle followed a specific Lagrangian trajectory, we revealed each trajectory using a fourth order Runge-Kutta interpolator. To obtain smooth Lagrangian trajectories, the original 4D flow MRI data were refined both in time (by two times) and space (by eight times) using cubic interpolation.

By evaluating the Euclidean arch length for each 3D trajectory *i* (originating from **x**₀) for a certain time interval $\tau$, we defined the value of the Lagrangian descriptors as a function **M** ($x_0$, $t_0$, $\tau$). Notably, the function **M** was in a discrete form when using the 4D flow MRI data; therefore, it will be denoted hereafter as **DM**. As the LCSs are defined by revealing the attracting and repelling structures, the evaluation of **DM** was performed both forward (**DM$^F$**) using $\tau_F$ and backward (**DM$^B$**) using $\tau_B$ in time to reveal the repelling and attracting structures, respectively.

$$DM_i = DM_i^F + DM_i^B = \sum_{n=-N_B}^{N_F} \sqrt{(x_i^{n+1} - x_i^n)^2 + (y_i^{n+1} - y_i^n)^2 + (z_i^{n+1} - z_i^n)^2} \quad (1.1)$$

$$DM_i^F = \sum_{n=0}^{N_F} \sqrt{(x_i^{n+1} - x_i^n)^2 + (y_i^{n+1} - y_i^n)^2 + (z_i^{n+1} - z_i^n)^2} \quad (1.2)$$

$$DM_i^B = \sum_{n=-N_B}^{0} \sqrt{(x_i^{n+1} - x_i^n)^2 + (y_i^{n+1} - y_i^n)^2 + (z_i^{n+1} - z_i^n)^2} \quad (1.3)$$

To extract the location of the 3D LCS from the **DM** function, a gradient filter was utilized. A detailed validation of the approach applied to a fundamental Arnold–Beltrami–Childress (ABC) case can be found in the supplemntary material.

Figure 1 (multimedia view) displays the velocity fields of intraventricular diastolic and systolic flows for a healthy patient (Panels 1 and 2) and a patient with severe aortic valve regurgitation (Panels 3 and 4) at two selected instants on a plane crossing both aortic and mitral valves. In a patient with a healthy aortic valve, a single jet emerges from the mitral valve towards the apex of the left ventricle during diastole (Panel 1). This inflow jet is expected to be directed smoothly towards the aortic valve to be ready for ventricular ejection during systole (Panel 2).

In a patient with aortic valve regurgitation, in addition to the jet emerging from the mitral valve, a jet also emerges from the aortic valve and is directed towards the mitral valve (Panel 3). This jet interacts with the mitral valve inflow and has been shown to result in kinematic obstruction of the flow and elevated viscous energy dissipation[35,36]. No significant differences with the healthy case can be observed on this plane during systole (Panel 4).

Figure 2 (multimedia view) shows the interventricular backward LCS in a healthy subject during early ventricular diastole (at T* =0.1, where T* = t/cardiac cycle time). To depict the LCS, three orthogonal slices are used to show the propagation of the coherent structures (in panel 4).

For practical reasons and to ease the visualization of the time evolution of LCS over a cardiac period, for the rest of this contribution, we will focus on one selected ventricular plane.

On the selected plane, three-time instants during early ventricular diastole (E-wave), late ventricular diastole (A-wave), and ventricular systole are reported. For the healthy case (Figure 3), during early diastole (T* = 0.1), the mitral jet structures are well-defined, including the jet front (indicated by the yellow arrow) and its deformed vortex ring (indicated by the green arrows) as shown in (Figure 3; Panel 1). The identified deformed vortex ring matches the reported intraventricular structures in [5,37,38] where a distinct asymmetric vortex ring was reported. As the cardiac cycle progresses, the formed structures lose the clear vortex structures, similar to the findings in[5], and are directed towards the aortic side (Figure 3; Panel 2). Moreover, the structures formed by the A-wave further direct the flow towards ejection, as noticed in Figure 3, Panel 2. This is similar to the finding reported in [38]. During systole (Figure 3; Panel 3), ridges of LCS are highly localized close to the aortic side, indicating the rapid arrangement of the fluid towards ejection.

For aortic regurgitation, Figure 4 shows the backward LCS structures on the plane crossing the mitral and aortic valves. More complex flow structures can be observed. During the E-wave, two distinct jets can be observed (indicated by the green and yellow arrows on Figure 4; Panel 1). The yellow arrow front is following the regular path found in healthy patients during the E-wave, while the green arrow front follows a transverse out-of-plane path directed towards the mitral valve. The interaction between the two jets leads to more complex flow structures and more active LCS in the left ventricle during the A-wave (Figure 4; Panel 2) and throughout the diastole. During systole (Figure 4; Panel 3), the ejected flow structures can be identified; however, a large region of the LV (towards the apex) still has actively moving LCS. This clearly indicates that the interventricular blood transport mechanism has been significantly altered due to the presence of aortic regurgitation.

As previously indicated, aortic regurgitation induces more complex flow structures in the left ventricle. To better visualize such structures, additional planes are depicted in Figure 5 (multimedia view). The selected time instant corresponds to T* = 0.29, or during early ventricular diastole (the E-wave). By overlaying the three-dimensional flow field and rotating the view to three different planes, it is possible to better highlight the specific out-of-plane flow structures induced by the regurgitant jet and its interaction with the mitral jet. In Figure 5 (multimedia view) (Panel 2), the E-wave jet is protruding through the plane corresponding to Panel (2), while Panel (3) captures the regurgitant jet following a semi-circular arch where it leaves the plane close to the aortic side and re-enters close to the mitral side. Clearly, to extract the LCS associated with such complex flow transport, a different plane needs to be considered, which is indicated by the blue-colored edge plane on Panel (4). A 2D view on this new plane shows the LCS corresponding to the E-wave at $T^* = 0.29$ (Panel 5) and $T^* = 0.39$ (Panel 6). These planes clearly capture the impingement between the jet and the LV wall besides showing the formation of the rolling vortical structures. The revealed structures show a complex fluid transport where it can be challenging to find a plane of symmetry for this transport. This advocates for the need to extract three-dimensional LCS to better understand the mechanism of blood transport, particularly in the presence of pathological conditions.

In summary, this study uses a novel approach to extract the interventricular LCS using 4D flow MRI data. We can list at least three major contributions of this study compared to previous studies [5,35]: 1) In this study, LCSs are visualized using the Lagrangian descriptors approach, which is computationally inexpensive compared to the classical approach based on finite-time Lyapunov exponents; 2) The extracted LCS rely on the full three-dimensional stretching experienced by each advected particle, while in previous studies, the reported LCSs were extracted solely based on a 2D representation only considering the two-dimensional stretching experienced by advected particles; 3) This is the first study extracting three-dimensional LCS in healthy patients and in patients with aortic regurgitation.

The main limitations associated with this study are the small population size (8 healthy controls and three patients with severe aortic regurgitation) and the fact that our cohort only includes patients with severe aortic regurgitation.

In this study, three-dimensional Lagrangian coherent structures have been extracted, for the first time, in healthy patients and in patients with aortic regurgitation. To do so, an approach based on Lagrangian descriptors has been used since it is more computationally efficient compared to probabilistic and geometric approaches. This study highlights the complex flow structures induced in the left ventricle in the presence of aortic regurgitation and the need to extract three-dimensional Lagrangian coherent structures to gain a better understanding of the complex interaction between the mitral inflow and the regurgitant jet.

# Declarations

**-Ethics approval and consent to participate**

**-Consent for publication**

Not applicable

**-Data Availability**

The data that support the findings of this study are available from the corresponding author upon reasonable request.

**-Competing interests**

The authors have no conflicts to disclose.

# Funding

LK acknowledges the support of the Natural Science and Engineering Research Council of Canada/Conseil de Recherche en Science Naturelles et en Génie du Canada (RGPIN-344164-07)

JG was funded by The University of Calgary, a start-up funding. We acknowledge the support of the Natural Science and Engineering Research Council of Canada/Conseil de Recherche en Science Naturelles et en Génie du Canada, RGPIN-2020-04549, and DGECR-2020-00204.

**-Authors' contributions**

LK and JG conceived the study. JG acquired and organized the raw data. AD developed the post-processing codes for LD computations. WA and AD post-processed the data and extracted the results. All authors participated in study design. All authors participated in revising the manuscript and read and approved the final manuscript.

# FIGURE 1

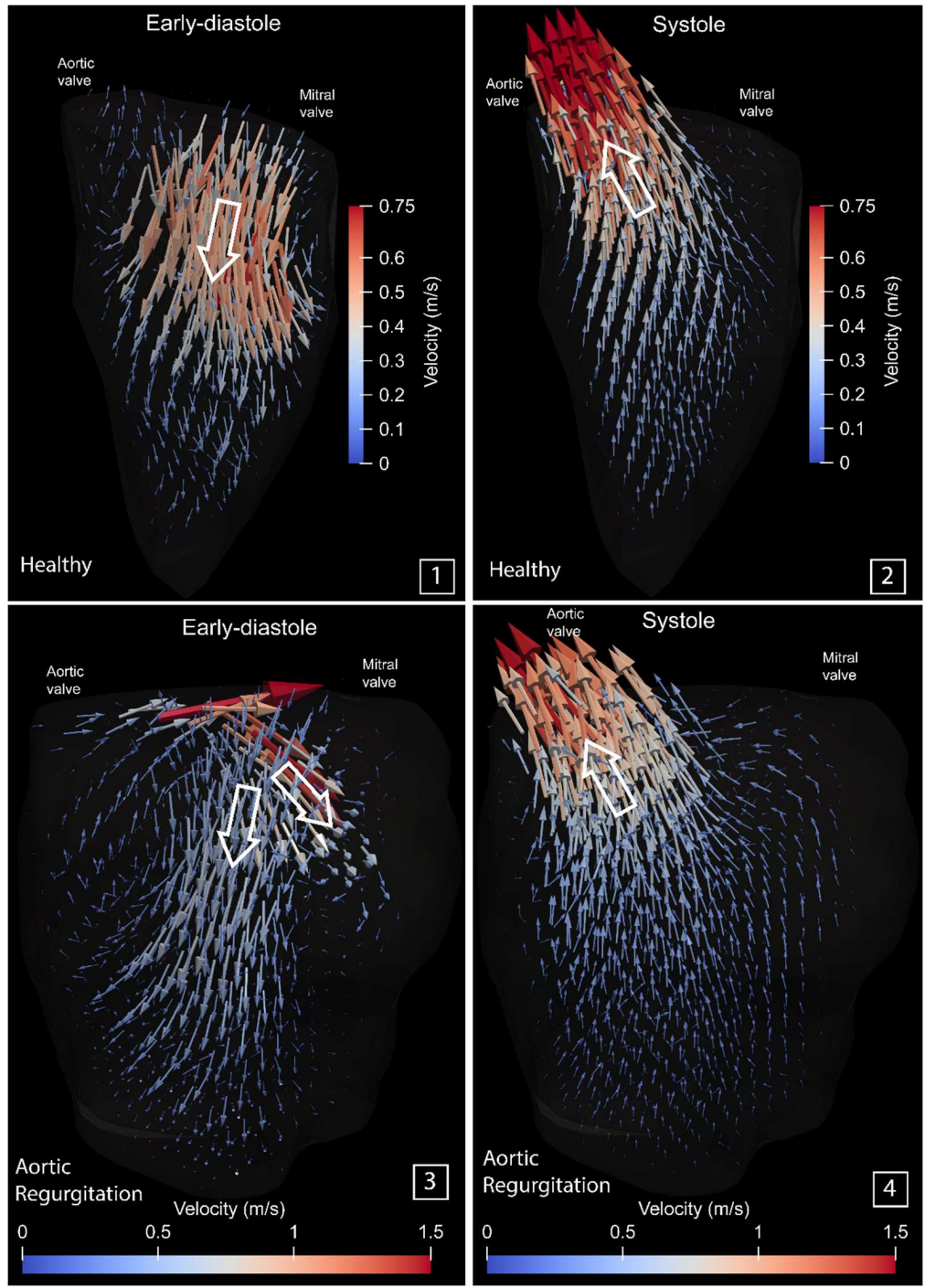

**Figure 1.** Three-dimensional velocity field on a selected left ventricle plane during systole and diastole for a healthy patient (Panels 1 and 2) and a patient with severe aortic regurgitation (Panels 3 and 4) (multimedia view) (Multimedia view).

# FIGURE 2

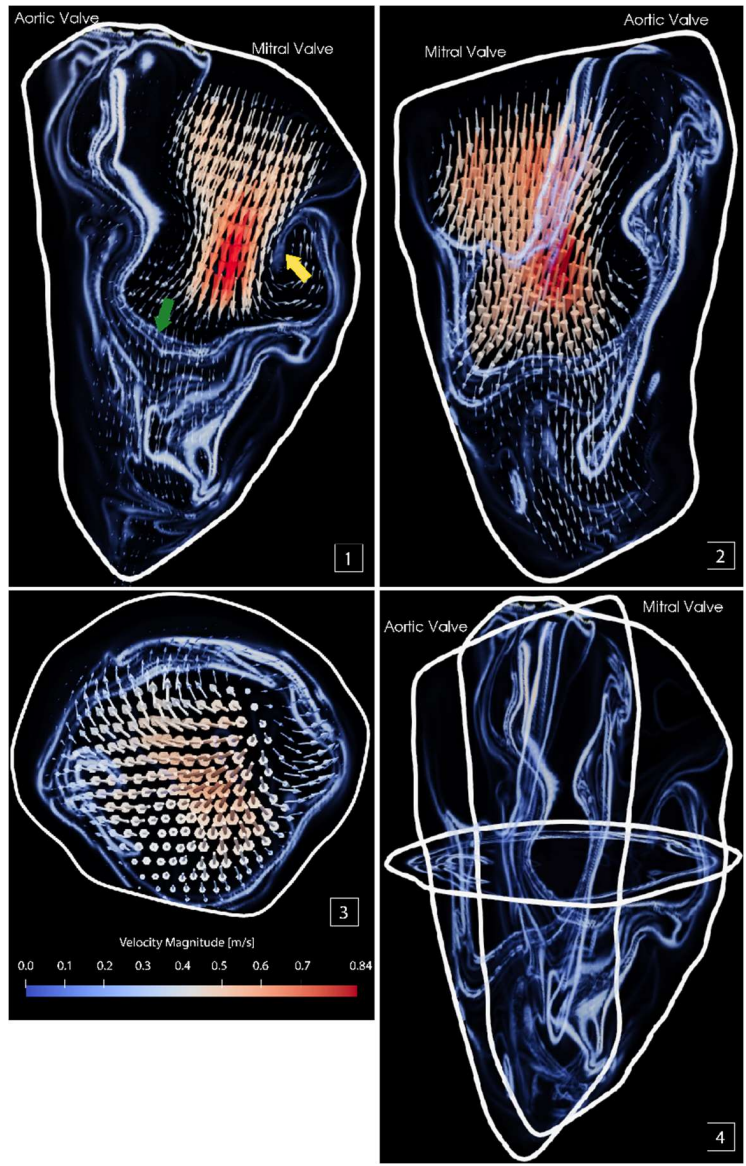

**Figure 2:** A three-dimensional representation of the backward Lagrangian coherent structures (LCS) for a healthy subject during the E-wave (T* = 0.1). Panels 1, 2, and 3 show the normal view of the three planes located in panel 4. In each view, the corresponding velocity field is extracted and depicted on the slice. Green and yellow arrows indicate some characteristic LCS commonly found in healthy patients during the filling phase (Green: jet front; Yellow: vortex ring) (multimedia view).

# FIGURE 3

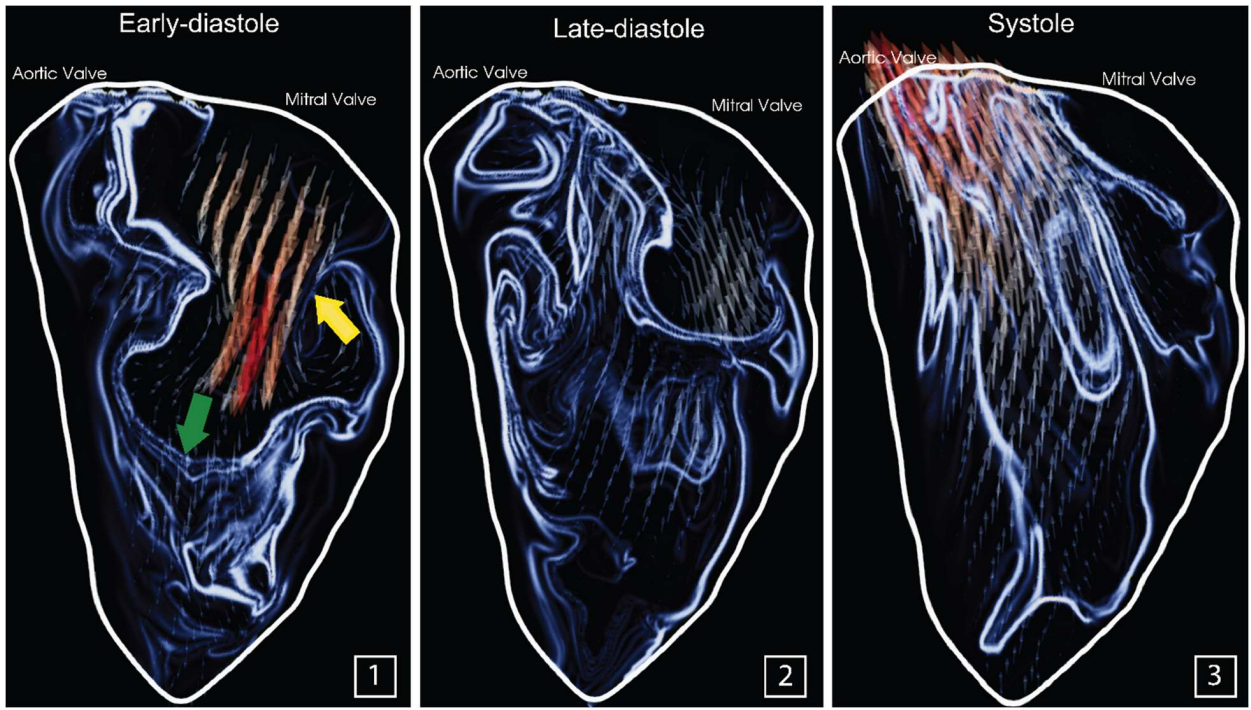

**Figure 3:** Backward Lagrangian coherent structures identified by the edges of backward Discrete M function of the Lagrangian descriptors in the left ventricle of a healthy control. 1) during E-wave; 2) during A-wave; 3) during systole.

# FIGURE 4

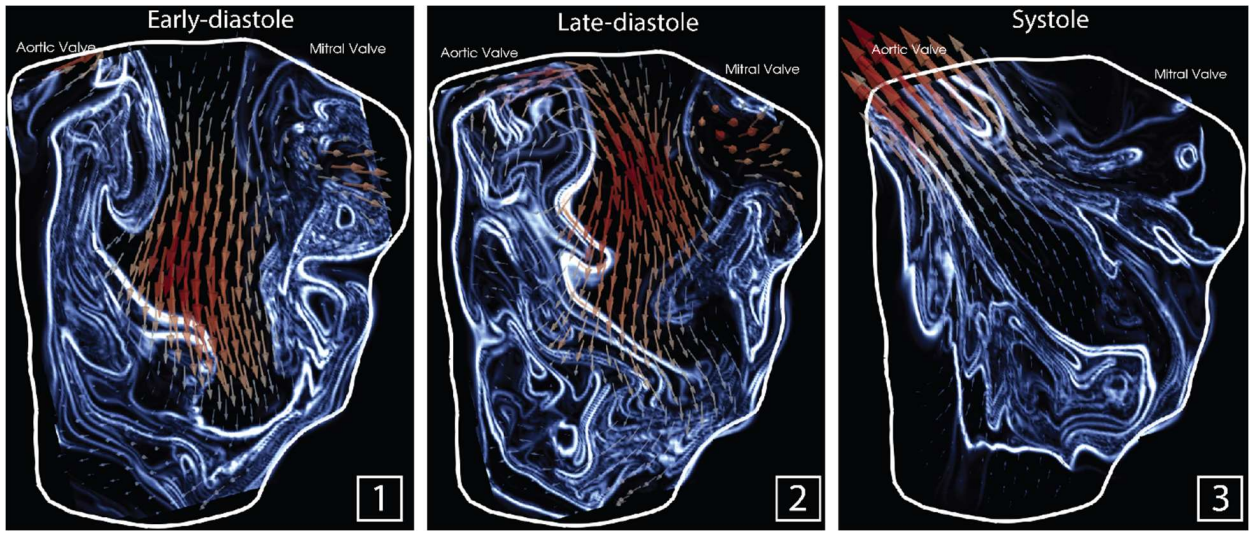

**Figure 4:** Backward Lagrangian coherent structures identified by the edges of backward Discrete M function of the Lagrangian descriptors in the left ventricle of a patient with severe aortic regurgitation. 1) during E-wave; 2) during A-wave; 3) during systole.

# FIGURE 5

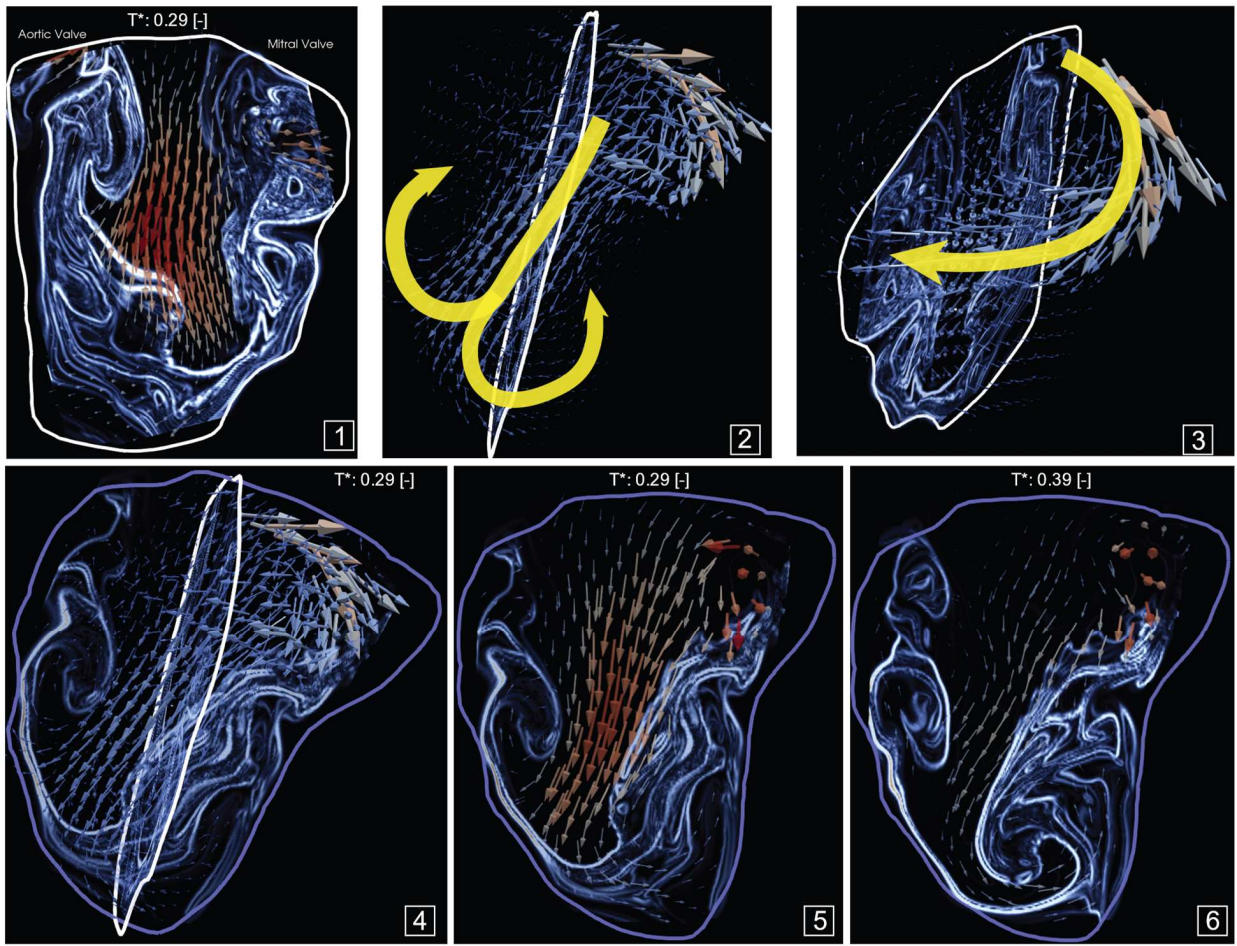

**Figure 5:** The complexity of backward LCS being revealed by showing additional planes. Panel 1 shows the 2D slice of backward LCS on the standard plane crossing both aortic and mitral valves. Panels 2 and 3 depict the same slice as in A along with the volumetric velocity field. Two directions for out-of-plane transport of the fluid are represented by yellow arrows. Plane 4 is selected to capture the travel of the mitral jet as marked in Panel 2. Planes 5 and 6 show the resulting structures by the interaction between the mitral jet and the left ventricular wall (multimedia view).